\documentclass[twocolumn,aps,prl,showpacs,groupedaddress,amsmath,amssymb]{revtex4}
\usepackage{amsmath}
\usepackage{amssymb}
\usepackage{txfonts}
\usepackage{graphicx}
\usepackage{dcolumn}
\begin{document}
\title{Proposal for a Loophole-Free Bell Test with Electron Spins of Donors}
\author{Fang-Yu Hong}
\author{Shi-Jie Xiong }

\affiliation{National Laboratory of Solid State Microstructures and
Department of Physics, Nanjing University, Nanjing 210093, China}
\date{\today}
\begin{abstract}
So far, all experimental tests of Bell inequalities which must be
satisfied by all local realistic hidden-variable theories and are
violated by quantum mechanical predictions have left at least one
loophole open. We propose a feasible setup allowing for a
loophole-free test of the Bell inequalities. Two electron spin
qubits of donors $^{31}P$ in semiconductors in different cavities
300 m apart are entangled through a bright coherent light and
postselections using homodyne measurements. The electron spins are
then read out randomly and independently by Alice and Bob,
respectively, with unity efficiency in less than 0.7$\mu$s by using
optically induced spin to charge transduction detected by
radio-frequency single electron transistor. A violation of Bell
inequality larger than 37\% and 18\% is achievable provided that the
detection accuracy is 0.99 and 0.95, respectively.

\end{abstract}


\maketitle

Most working scientists hold fast to the concepts of 'realism'
according to which an external reality ezists independent of
observation and 'locality' which means that local events cannot be
affected by actions in space-like separated regions \cite{jfcs}. The
significance of these concepts goes far beyond  science.   Based on
these  deep-rooted reasonable assumptions, in their seminal 1935
paper, Einstein, Podolsky, and Rosen (EPR) advocated  that quantum
mechanics is incomplete \cite{aepb}. The EPR arguments about the
physical reality of quantum systems is shifted from the realm of
philosophy to the domain of experimental physics  since 1964 when
Bell and others constructed mathematical inequalities - one of the
profound scientific discoveries of the 20th century \cite{jsb,jfch},
which must be satisfied by any theory based on the joint assumption
of realism and locality and be violated by quantum mechanics. Many
experiments \cite{jfcs,sjfc,esft,aagp,zyom,prtr,pgkm,wtbj,mard} have
since been done that are consistent with quantum mechanics and
inconsistent with local realism. So far, however, all these tests
suffered from "loopholes" allowing a local-realistic explanation of
the experimental results by exploiting either the low detector
efficiency \cite{pmre,esan} or the timelike interval between the
detection events \cite{jbel,sant}. The first loophole names the
detection loophole allowing the possibility that the subensemble of
detected events agrees with quantum mechanics even though the entire
ensemble is consistent with Bell inequalities. So a fair-sampling
hypothesis that the detected events represent the entire ensemble
must be assumed. The second refers the locality or 'lightcone'
loophole allowing the correlations of apparently separate events
resulting from unknown subluminal signals which propagate between
space-like regions of the apparatus to take place.

Several schemes were proposed closing these loopholes based on
entangled photon pairs \cite{pgkw,sfhm}, Hg atoms \cite{esfw},
Rydberg atoms \cite{mfre}, trapped ions \cite{cswi}, or non-Gaussian
states of light and balanced homodyning \cite{hnhc,rgpf}, but all
face a formidable experimental challenge.  Here we propose a scheme
for the loophole-free Bell test based on the Kane Si:P architecture
\cite{beka}, in which two qubits are encoded onto two electron spins
of donor atoms $^{31}\text{P}$ in doped silicon electronic devices
in two high-Q cavities  300 m apart. Entanglement between the two
qubits is created by using bright coherent-light pulses which
interact with the donor atoms $^{31}\text{P}$ through a weak
dispersive light-matter interaction, respectively, via homodyne
detections and postselections \cite{pvll}. The qubits are then read
out with perfect efficiency and high accuracy above 99\% in about
0.6 $\mu$s using optically induced spin to charge transduction
\cite{rjsw,tmbr,mjtg}. The realization of the setup is within the
ability of the current semiconductor fabrication technology
\cite{mlbd,dbor,mksw}, and the read-out of the single donor electron
spin can be realized through resonant spin-dependent charge transfer
where the resulting electron current is measurable using
radio-frequency single electron transistor (rf-SET)
\cite{beka,rjsw,tmbr,mjtg,lchw}.

A Bell measurement of inequality of Clauser, Horne, Shimony, and
Holt (CHSH) \cite{jfch} comprises of three basic ingredients. First,
a pair of particles entangled with each other is prepared in a
repeatable starting configuration. Then a variable classical
manipulation is applied independently and randomly to each particle;
these manipulations are labeled as $\phi_1$ and $\phi_2$. At last, a
classical property with two possible outcome values 1 and -1 is
measured for each of the particles. The correlation is measured by
repeating the experiment many times and can be expressed as
\begin{equation}\label{eq1}
q(\phi_1,\phi_2)=\frac{N_s(\phi_1,\phi_2)-N_d(\phi_1,\phi_2)}{N_s+N_d},
\end{equation}
where $N_s$ and $N_d$ denote the numbers of measurements in which
the measured results are the same or different, respectively. The
CHSH form of Bell inequalities states that the correlations
resulting from local realistic theories must satisfy:
\begin{equation}
\begin{split}\label{eq2}
S(\alpha_1,\delta_1,\beta_2,\gamma_2)&=|q(\alpha_1,\beta_2)-q(\alpha_1,\gamma_2)\\
&+q(\delta_1,\beta_2)+q(\delta_1,\gamma_2)|\leq2,
\end{split}
\end{equation}
where $\alpha_1$ and $\delta_1$ ( $\beta_2$ and $\gamma_2$ ) are
specific values of $\phi_1$ ($\phi_2$). For a Bell measurement based
on  electron spins, we have
\begin{equation}
q(\phi_1,\phi_2)=\langle\psi|(\boldsymbol{\sigma} ^A\cdot
\textbf{n}_1)(\boldsymbol{\sigma}^B\cdot \textbf{n}_2)|\psi\rangle,
\end{equation}
where $\boldsymbol{\sigma}\,^i=(\sigma_x\,^i, \sigma_y\,^i,
\sigma_z\,^i)$ with $\sigma_j\,^i$ ($i=A, B$, and $j=x, y, z$) being
the Pauli matrices, and $\textbf{n}\,_i$ ($i=1, 2$) are unit
vectors. The CHSH inequality (\ref{eq2}) is maximally violated by
quantum
 mechanics at certain sets of $\textbf{n}_1$ and $ \textbf{n}_2 $, one
such set is that both of the $\textbf{n}\,_i$ ( i=1, 2 ) are in the
$xy$-plane, and the polar angles of $\textbf{n}\,_i$ ( i=1, 2 ) are
$\alpha_1=0$, $\delta_1=\pi/2$ for $\textbf{n}_1$, and
$\beta_2=\pi/4$, $\gamma_2=3\pi/4$ for $\textbf{n}_2$. For these
phase angles and state
$|\psi\rangle=(|\downarrow\uparrow\rangle-|\uparrow\downarrow\rangle)/\sqrt2$,
the quantum mechanics gives
\begin{equation}
B\left(0,\frac{\pi}{2},\frac{\pi}{4},\frac{3\pi}{4}\right)=S\left(0,\frac{\pi}{2},
\frac{\pi}{4},\frac{3\pi}{4}\right)-2=2\sqrt2-2.
\end{equation}

\begin{figure}
\includegraphics[scale=0.35]{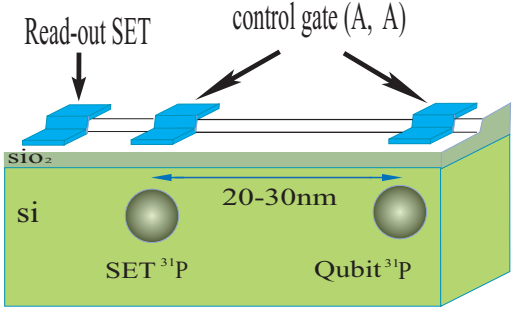}
\caption{\label{fig:1} (Color online) Schematic picture of the
device for the resonant spin-dependent charge transfer of a single
electron of a donor $^{31}P$ in silicon. The tunneling event
$D^0D^0\rightarrow D^+D^-$ is Pauli blocked if the qubit donor and
the SET-donor spins are parallel.}
\end{figure}

The architecture of the basic phosphorus $^{31}\text{P}$ donor
electron spin qubit in silicon with control gates and a resonant
readout mechanism are shown in Fig. \ref{fig:1}. The donors serve to
localize the electron spins in space which encode quantum
information in the conventional fashion as
$|0\rangle=|\downarrow\rangle=(0,1)^\dag$ and
$|1\rangle=|\uparrow\rangle=(1,0)^\dag$, and to provide local qubit
addressability through the electron-nuclear hyperfine interaction
\cite{beka}. Indirect spin detection involves transfer of the spin
information to the charge degrees of freedom through a
spin-dependent tunneling process, during which the resulting
electron current can be detected by an ultrasensitive electrometer,
rf-SET \cite{rjsw}. This concept depends on the application of a
small dc electric field $F_{dc}$ and an ac electric field $F_{ac}$
with the amplitude $F_{dc}\ll F_{ac}$ resonant at the energy gap
$\Delta E$ of the two states $D^0$ and $D^-$ to induce the tunneling
of the qubit electron to a secondary (spin polarized) ``SET-donor".
Here $D^-$ denotes the state with two electrons being bound to the
same donor formed by the tunneling. The resulting charge
re-distribution can be detected by a rf-SET. This detection is
equivalent to the measurement of the qubit spin state, because the
tunneling event $D^0D^0\rightarrow D^+D^-$ will be Pauli blocked if
the qubit and the SET-donor electron spins are parallel.

To generate entanglement between donor qubits at Alice's and Bob's
sites  300 m apart, a bright coherent pulse sequentially interacts
with the qubits, entangled qubit pairs will then be postselected
conditioned upon the results of probe homodyne measurements. For a
sufficient dispersive interaction between the donor electron and the
light, the system should be placed in a cavity resonant with the
light. For the cavity, weak coupling is sufficient, but a high value
of $Q/V$ is required, where $Q$ is the quality and $V$ is the
mode-volume of the cavity \cite{pvll}.

The donor electron spin system in the cavity is treated as a
$\Lambda$ system with two stable and metastable ground states
$|0\rangle$ and $|1\rangle$, and an excited state $|e\rangle$ provided  by the
 bound-exciton state. For the coherent pulses, the transition between $|0\rangle$ and $|e\rangle$ is suppressed due to a prohibitive selection rule  and only $|1\rangle$ and $|e\rangle$ participate in the
interaction with the cavity mode \cite{tdll}. Assuming that the state of the
qubits in Alice's and Bob's sites are initially prepared in the
states $(|0\rangle-|1\rangle)/\sqrt2$ and
$(|0\rangle+|1\rangle)/\sqrt2$, respectively, and the two SET $^{31}
P$ are in $|0\rangle$ state that makes them not to take part in the
interaction with the bright coherent pulses. The coherent light is
sufficiently detuned from the transition between $|1\rangle$ and the
excited state to allow for a strictly dispersive light-matter
interaction. When the probe pulse in coherent state $|\alpha\rangle$
reflects from the cavity at the Alice's site, the total output state
may be described by \cite{pvll}
\begin{equation}
\hat{U}_{int}[(|0\rangle-|1\rangle)
|\alpha\rangle]/\sqrt2=(|0\rangle|\alpha\rangle-|1\rangle|\alpha
e^{-i\theta}\rangle)/\sqrt2.
\end{equation}
The probe beam is then sent to the cavity at the Bob's site and
interacts with the qubit donor in the same way. Applying a further
linear phase shift of $\theta$ to the pulse after it leaves the
cavity will yield the total state
\begin{equation}
|\psi\rangle=\frac{1}{2}(\sqrt2|\psi^-\rangle|\alpha\rangle+|00\rangle|\alpha
e^{i\theta}\rangle-|11\rangle|\alpha e^{-i\theta}\rangle),
\end{equation}
where $|\psi^-\rangle=(|01\rangle-|10\rangle)/\sqrt2$ with the
conventional denotation $|01\rangle=|0\rangle_A|1\rangle_B$ and
$\langle01|=\,_A\langle0|_B\langle1|$ hereafter.

In the presence of channel loss, we may model the photon loss by
considering a beam splitter in the channel that transmits only a
part of the probe pulse with transmission $\eta^2$ \cite{pvll}. Tracing over the
losed photons introduces  decoherence and the total state can be
described by density matrix $\rho$, with the following diagonal part
for the pulse states
\begin{equation}
\begin{split}
 \rho_{dia}=&\frac{1}{4}|00\rangle\langle00|\eta\alpha
e^{i\theta}\rangle\langle\eta \alpha
e^{i\theta}|-\frac{1}{4}|11\rangle\langle11|\eta\alpha
e^{-i\theta}\rangle\langle \eta\alpha
e^{-i\theta}|\\&+\frac{1}{2}\rho_{en}|\eta\alpha \rangle\langle\eta \alpha
|,
\end{split}
\end{equation}
where
\begin{equation}
\rho_{en}=\frac{1}{2}|01\rangle\langle01|-|01\rangle\langle10|e^{-\gamma+i\xi}\zeta-
|10\rangle\langle01|e^{-\gamma-i\xi}\zeta+|10\rangle\langle10|.
\end{equation}
Here $\zeta$ is the decoherence factor arising from the dispersive
light-matter interaction in the cavities,
$\gamma\approx\frac{1}{2}(1-\eta^2)\alpha^2\sin^2\theta=\frac{1}{2}(1-\eta^2)d^2$,
and an extra phase $\xi=\alpha^2(1-\eta^2)\sin\theta$  can be set to
be naught, since it is independent of the measurement results and
can be locally removed via static phase shifters.

With the balanced homodyne detection \cite{scu}, the success
probability of generating entanglement between two qubits at Alice's
and Bob's sites is found to be \cite{pvll}
\begin{equation}\label{eq14}
P_s=\text{Tr}\int_{-p_c}^{+p_c}\rho\,
dp=\frac{\text{erf}(b_0)}{2}+\frac{\text{erf}(b_1)}{4}+\frac{\text{erf}(b_{-1})}{4},
\end{equation}
where $b_s=\sqrt2(p_c+s\eta d)$, $s=0,\pm1$, and $p_c$ is the
selection window of the homodyne measurements. The desired entangled
output state is $|\psi^-\rangle$,
 so the average fidelity
after postselection has the form \cite{pvll}
\begin{equation}\label{eq15}
\begin{split}
F&=\frac{1}{P_s}\left[\int_{-p_c}^{+p_c}dp\langle\psi^-|\rho|\psi^-\rangle\right]\\
 &=\frac{\text{erf}(b_0)(1+e^{-\gamma})}{2\text{erf}(b_0)+\text{erf}(b_1)+\text{erf}(b_{-1})}.
\end{split}
\end{equation}
For the state obtained through the postselection with the
configuration for $\textbf{n}_i$ (i=1,2) aforesaid, the violation of
the CHSH inequalities reads
 \begin{equation}\label{eq3}
B(0,\frac{\pi}{2},\frac{\pi}{4},
\frac{3\pi}{4})=2\sqrt2e^{-\gamma}\zeta\frac{\text{erf}(b_0)}{2P_s}-2.
 \end{equation}
After the pulse leaves the cavity at the Bob's site, Alice and Bob
randomly and dependently manipulate the electron spins of the SET
$^{31}P$ from the initial state $|0\rangle$ to the state
$(|0\rangle+|1\rangle)/\sqrt2$ corresponding to $\alpha_1=0$ or the
state $(|1\rangle+e^{i\frac{\pi}{2}}|0\rangle)/\sqrt2$
($\delta_1=\frac{\pi}{2}$) and to the state
$(|1\rangle+e^{i\frac{\pi}{4}}|0\rangle)/\sqrt2$ (
$\beta_2=\frac{\pi}{4}$) or the state
$(|1\rangle+e^{i\frac{3\pi}{4}}|0\rangle)/\sqrt2$
($\gamma_2=\frac{3\pi}{4}$), respectively. These manipulations on
the SET $^{31}P$ electron spins equivalent to the actions on the
qubit donor spins can be finished in 0.1 $\mu$s \cite{cdhh}. Then
Alice and Bob read out the electron spins of the qubit $^{31}P$
using optically induced spin to charge transduction detected by the
rf-SET. If the qubit and SET-donor electron spins are parallel, the
tunneling event $D^0D^0\rightarrow D^+D^-$ will be Pauli blocked,
thus the rf-SET will detect nothing, that will be assigned a value
+1. Otherwise, the rf-SET will detect a current signal, and the
outcome will be assigned -1. The outcomes of every experiment can be
used to compute the correlation function $q(\phi_1,\phi_2)$, so that
fair-sampling hypothesis is not required. In this case, the issue of
the detection efficiency is replaced by the detection accuracy
$\kappa$. The main origin of the inaccuracy in the detection of
electron spins comes from the imperfection in the spin to charge
transduction, i.e., the tunneling event $D^0D^0\rightarrow D^+D^-$
may not happen even if the qubit and SET-donor electron spins are
antiparallel. This leads to an experimental result that the
probability of the qubit spin parallel to the SET-donor spin will be
larger than that obtained with the perfect detection accuracy. The
read-out of the qubit spin with the detection accuracy
$\kappa\geq0.99$ may be possible according to \cite{mjtg,lchw}.
Considering the detection accuracy $\kappa$ and the detection error
rate $\tau=1-\kappa$, the violation of the CHSH inequalities $B$ in
Eq. (\ref{eq3}) may be rewritten as
\begin{equation}\label{eq4}
B(0,\frac{\pi}{2},\frac{\pi}{4},\frac{3\pi}{4})=
2\sqrt2e^{-\gamma}\zeta(1-\tau)^2 \frac{\text{erf}(b_0)}{2P_s}
-2\tau^2-2.
 \end{equation}

\begin{figure}
\includegraphics[scale=0.45]{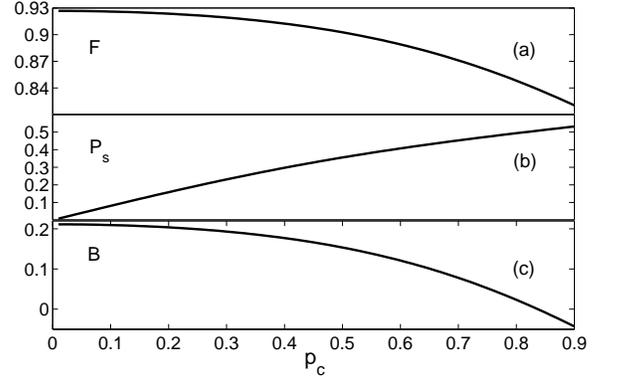}
\caption{\label{fig:2} The fidelity of the obtained state by
postselection (a), the success probability $P_s$ (b), and the
violation of the CHSH inequalities
$B\left(0,\frac{\pi}{2},\frac{\pi}{4},\frac{3\pi}{4}\right)$ (c) as
functions of the postselection window $p_c$ in the case of the
detection accuracy $\kappa=0.95$. See the text for the values of
other parameters.}
\end{figure}

\begin{figure}
\includegraphics[scale=0.45]{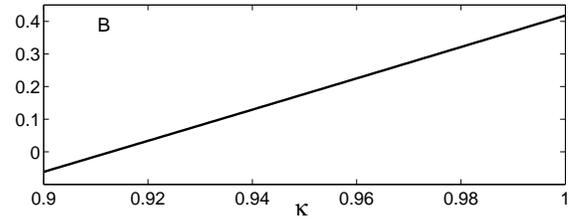}
\caption{\label{fig:3}(a) The violation of the CHSH inequalities
$B(0,\frac{\pi}{2},\frac{\pi}{4},\frac{3\pi}{4})$ versus the
detection accuracy $\kappa$ with $p_c=0.4$. See the text for the
values of other parameters.}
\end{figure}

Assuming the telecom fiber and wavelength where losses are about 1
dB/km \cite{nggr}, the transmission parameter of 300 m is
$\eta^2=10^{-0.03}$. For $^{31}P$ donor impurity in silicon,
distinguishability of $d=1.5$ corresponding to $\alpha=100$ and
$\theta=0.015$ is achievable \cite{tdll}. Assuming that the
decoherence factor $\zeta=0.95$ and the detection accuracy
$\kappa=0.95$, from Eqs. (\ref{eq14}), (\ref{eq15}), and
(\ref{eq4}), we obtain the results shown in Fig. \ref{fig:2}.
Assuming that the selection window $p_c=0.4$ and other parameters
unchanged, we get the relation between the violation of the CHSH
inequalities $B$ and the detection accuracy $\kappa$ as shown in
Fig. \ref{fig:3}. When $p_c=0.4$ and $\kappa=0.99$, we have
$P_s=0.30$, $F=0.91$ and $B=0.37$. Even the detection accuracy is so
low that $\kappa=0.95$, we still have $B=0.18$ for $p_c=0.4$. With a
repetition rate of 500 kHz and $P_s=0.30$, the number of data
samples would be 150,000 per second, thus the whole Bell test
experiments would be finished in less than one second.

\begin{figure}
\includegraphics[scale=0.3]{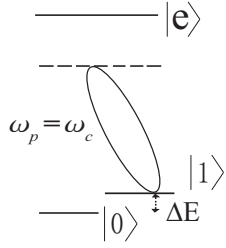}
\caption{\label{fig:4} The schematic of relevant energy structure of a phosphorous impurity in silicon, probe pulses, and the  cavity. The probe pulses are on resonant with the cavity, $\omega_p=\omega_c=0.75$ eV, and far detuned from the atomic transition $|1\rangle\rightarrow|e\rangle$. Energy split $\Delta E=1$ meV for an applied 8.6 T
    magnetic field. The energy difference between $|e\rangle$ and $|0\rangle$ is about 1.09 eV.}
\end{figure}

Building the  setup shown in Fig. \ref{fig:1} is within the reach of the
current technology. The two donors of distance $20\sim30$ nm can be placed
through random doping  techniques, though only a small percentage of such
devices will work properly \cite{beka}. The three gates  with lateral dimensions
 and separation $\sim 10$ nm can be patterned on the surface through
 technologies  such as self-assembly and  the use of extreme ultraviolet
  radiation, x-rays, and electron beams \cite{mlbd,gyle}. A workable transistor with a
  gate length of $6$ nm have already been realized in 2002 \cite{mlbd,bdor}.
   The two qubit states have a energy split $\Delta E=1$ meV for an applied 8.6 T
    magnetic field. The exited state $|e\rangle$ is about 1.09 eV
     above the ground state \cite{dkmt}. The probe coherent pulses of wavelength about 1650 nm
     ($\omega_p=0.75$ eV) (Fig. \ref{fig:4}) are far detuned from  the transition $|1\rangle\rightarrow|e\rangle$, but  on resonant with the cavity with frequency $\omega_c$, of which the device of dimensions about 50 nm$\times$30 nm$\times$30 nm  is at the antinode \cite{tdll}.

As a summary, we present a scheme for the loophole-free test of the
Bell inequalities. The detection efficiency of donor electron spins
is unity using the optically induced spin to charge transfer
detected by an rf-SET, and the fair sampling assumption is not
required, thereby the detection loophole in this scheme is closed.
The two qubit donors are 300 m apart, and the time of the random and
independent measurement of the two qubits by Alice and Bob,
respectively, is within 0.7 $\mu$s, thus the lightcone loophole may
be closed too. The experimental realization of this scheme is within
the reach of the current technology. Large violation of the CHSH
inequality $B=0.37$ for the detection accuracy $\kappa=0.99$ is
achievable. Even if the detection accuracy is so low that
$\kappa=0.95$, we may still have $B=0.18$. This scheme may open a
promising avenue towards a complete experimental Bell test which has
a profound significance far beyond science.

{\it Acknowledgments} This work was supported by the State Key
Programs for Basic Research of China (2005CB623605 and
2006CB921803), and by National Foundation of Natural Science in
China Grant Nos. 10474033 and 60676056.

\end{document}